\documentclass[
preprintnumbers,endnote,nofootinbib,superscriptaddress,
preprint]{revtex4}
\usepackage{graphicx}
\usepackage{amsmath}

\usepackage{color}

\usepackage[hypertex]{hyperref}
\newcommand{\vev}[1]{\langle {#1} \rangle}
\newcommand{\lsim}{\lesssim}
\newcommand{\gsim}{\gtrsim}

\newcommand{\ord}[1]{\mathcal{O}{\left(#1\right)}}
\newcommand{\beq}{\begin{equation}}
\newcommand{\eeq}{\end{equation}}

\newcommand{\mP}{\bar{M}_{\rm P}}
\newcommand{\Lphi}{\Lambda_\phi}
\newcommand{\ktil}{\tilde{k}}
\newcommand\invfb{\ensuremath{\text{fb}^{-1}}}
\newcommand{\mkk}{m_{\rm KK}}

\newcommand\ETmiss{\ensuremath{E_T^\text{miss}}}
\newcommand\p{\mathbf p}
\newcommand\pTmiss{\ensuremath{\p_T^\text{miss}}}
\newcommand\pTll{\ensuremath{\p_T^{ll}}}
\newcommand\mll{\ensuremath{m_{ll}}}

\newcommand\code{\textsc}

\begin{document}

\pagestyle{plain}

\title{Radion as a Harbinger of Deca-TeV Physics}

\author{Hooman Davoudiasl
\footnote{email: hooman@bnl.gov} } \affiliation{Department of
Physics, Brookhaven National Laboratory, Upton, NY 11973, USA}

\author{Thomas McElmurry
\footnote{email: mcelmurry@pas.rochester.edu}
}
\affiliation{Department of Physics and Astronomy,
University of Rochester, Rochester, NY 14627, USA}

\author{Amarjit Soni
\footnote{email: soni@bnl.gov} } \affiliation{Department of Physics,
Brookhaven National Laboratory, Upton, NY 11973, USA}


\begin{abstract}

Precision data generally require the threshold for physics beyond the Standard Model to be at 
the deca-TeV (10~TeV) scale or higher.  This raises the question of whether there are 
interesting deca-TeV models for which the LHC may find direct clues.  A possible 
scenario for such physics is a 5D warped model of fermion masses and mixing, 
with Kaluza-Klein masses $\mkk\sim 10$~TeV, allowing it to avoid 
tension with stringent constraints, especially from flavor data.  
Discovery of a Standard-Model-like Higgs boson, for
which there are some hints at $\sim 125$~GeV at the LHC, would also 
require the KK masses to be at or above 10 TeV.  These warped models generically 
predict the appearance of a much lighter {\it radion} scalar.    
We find that, in viable warped models of flavor, 
a radion with a mass of a few hundred GeV and an inverse 
coupling of order $\mkk \sim 10$~TeV could typically be 
accessible to the LHC experiments---with $\sqrt{s} = 14$~TeV and 
$\sim 100$~\invfb\ of data.
The above statements can be applied, {\it
mutatis mutandis},  to 4D dual models, where conformal dynamics and
a dilaton replace warping and the radion, respectively.  
Detection of such a light and narrow 
scalar could thus herald the proximity of a new physical threshold and motivate 
experiments that would directly probe the deca-TeV mass scale.

\end{abstract}
\maketitle

\section{Introduction} 

A main goal of experiments at the
Large Hadron Collider (LHC) is the discovery of the mechanism for
electroweak symmetry breaking (EWSB).  While EWSB can be realized in
a variety of ways in Nature, the most economical possibility is
through a Higgs doublet scalar $H$ with a vacuum expectation value
$\vev{H}\simeq 250$~GeV, as in the minimal Standard Model (SM).
Based on precision electroweak data, it  is widely expected that the
SM Higgs mass $m_H\lsim 160$~GeV \cite{LEPEWWG,GfitterSM}.  To avoid
violations of perturbative unitarity (the onset of strong
interactions), the Higgs cannot be too heavy: $m_H\lsim 1$~TeV
\cite{uh1,uh2}.

The ongoing searches at the LHC have roughly yielded, at $95\%$
confidence level, $115~{\rm GeV}\lsim m_H \lsim 130$~GeV or else $m_H\gsim
500$~GeV \cite{ATLAS-Higgs,CMS-Higgs}, as of the time of the writing
of this paper.  Currently, both ATLAS \cite{ATLAS-Higgs} and 
CMS \cite{CMS-Higgs} report excess events, at about the $2\sigma$ level, 
that are consistent with a SM-like Higgs boson with a mass of $m_H \simeq 125$~GeV 
{\cite{note-added}}.

If the light Higgs signal at the LHC persists, one is compelled to
think what new physics may help stabilize its mass against quadratic
divergences that lead to the well-known hierarchy problem.  An
interesting possibility for such new physics is provided by 5D
warped models of hierarchy and flavor, based on the Randall-Sundrum
(RS) geometry \cite{Randall:1999ee}. The original RS model was
introduced to address the hierarchy between $\vev{H}$ and the Planck
scale $\mP \sim 10^{18}$~GeV.  The inclusion of the SM gauge fields
\cite{Davoudiasl:1999tf,Pomarol:1999ad} and fermions
\cite{Grossman:1999ra} in the 5D RS bulk can result in a predictive
framework for explaining the hierarchy and flavor puzzles
simultaneously \cite{Grossman:1999ra,Gherghetta:2000qt}. A natural
expectation in this scenario is the emergence of various
Kaluza-Klein (KK) resonances at the TeV scale.

While the simultaneous resolution of Planck-weak hierarchy and flavor
puzzle that warped models offer is highly attractive, it entails
significant corrections to electroweak precision observables, in particular those related to
the oblique $T$ parameter, which result in constraining the
KK-particle masses to above $\sim 10$~TeV \cite{Carena:2003fx}.  
This of course means that there still remains a
small hierarchy requiring some degree of tuning of $\ord{10^{-3}}$. 
Compliance with these bounds without tuned parameters requires an
enlarged bulk gauge group to provide a 
custodial symmetry \cite{custodial1,custodial2}.  
With this added complexity, the KK scale can be lowered to about 3~TeV
\cite{custodial1,custodial2,Carena:2006bn,Carena:2007ua} and the required tuning
then becomes only around $10^{-2}$. However,  this setup  then
becomes considerably less economical, requiring 
extension from the $SU(2)_L$ gauge symmetry 
to $SU(2)_L \times SU(2)_R$  and an added set of new particles.  
Moreover,  these interesting  attempts end up facing further hurdles
from the flavor sector, especially as the $K$-$\bar K$ mixing data \cite{Bona:2007vi} still
constrain KK masses to be near or above 10 TeV \cite{fb1,fb2}, unless one
resorts to some tuning \cite{fb3,Blanke:2008zb} 
or some additional symmetries \cite{Bauer:2011ah}.  Therefore, by accepting a fine-tuning of 
$\ord{10^{-3}}$, corresponding to KK masses of order 10~TeV, 
one retains the attractive simplicity of the warped models that address SM flavor.

It has been pointed out in Refs.~\cite{Azatov:2010pf,Goertz:2011hj,Carena:2012fk} that
if the Higgs properties are established to be close to those in
the SM, KK masses in warped models (with or without custodial symmetries)
are pushed to scales of order 10~TeV, well beyond the reach of the LHC \cite{LHCKK}.  Hence, the
confirmation of a SM-like Higgs state at the LHC would typically
constrain $\mkk$ to be well above the TeV scale, regardless of other
precision data.  Here, we note that while in Ref.~\cite{Azatov:2010pf}
the Higgs signal is predicted to be enhanced by the effects of the warped KK states, 
Refs.~\cite{Goertz:2011hj,Carena:2012fk} arrive at the opposite conclusion, namely a suppressed Higgs
signal.  The analysis in Ref.~\cite{Carena:2012fk} ascribes this discrepancy to the difference
in the regularization methods employed by the authors of Ref.~\cite{Azatov:2010pf} in
reaching their conclusions.  We do not comment here on which procedure may be the correct
approach.  However, either way, it is clear that the effects of warped states
would require a high KK mass scale, near 10~TeV, if significant departures
from SM predictions for the Higgs production and decay rates are not
detected at the LHC \cite{EarlyHiggs}.

The above considerations suggest that the simplest warped models of
hierarchy and flavor, especially those with a SM-like Higgs, would
be naturally characterized by values of $\mkk$ that lie outside the
reach of the LHC. For example, if KK states are at the deca-TeV
(10~TeV) scale, a simple and compelling picture of flavor can be
obtained that can comply with the most severe flavor constraints,
given the built-in RS Glashow-Iliopoulos-Maiani mechanism in these models 
\cite{Agashe:2004ay,Agashe:2004cp}. Without a custodial symmetry,
typically deviations from the precision bounds on the $T$ parameter
arise, albeit at modest levels for such a large KK mass scale.
Therefore, if the Higgs turns out to be light, with $m_H\sim
125$~GeV, new deca-TeV physics may need a mild degree of custodial protection. 
However, without access to KK modes at the LHC, it may
appear that we have achieved freedom from tension with flavor and
electroweak constraints at the expense of experimental
verifiability. We will argue below that this is not necessarily the
case.

In this work, we note that deca-TeV warped scenarios typically
include a light scalar, namely the radion $\phi$ of mass $m_\phi\ll
\mkk$, that may be accessible to TeV-scale experiments, such as those at the LHC.  The
appearance of such a scalar, often referred to as the dilaton,  is
also likely common to all dynamical EWSB theories that are
holographically dual \cite{ADSCFT} to a warped model
\cite{holography}, {\it i.e.} 4D models that are characterized by
conformal behavior above the KK scale
\cite{Goldberger:2007zk,Fan:2008jk,Vecchi:2010gj,Appelquist:2010gy}.
The couplings of $\phi$ are suppressed by the scale of new dynamics 
(mass scale of heavy resonances). If measured,  
the signal rate in various decay channels of $\phi$ and its narrow width 
could provide estimates of the scale that suppresses the 
interactions of $\phi$, offering clues about a 
new physical threshold near the deca-TeV scale.  We note that the width of the radion 
in the regime studied in our paper is typically much smaller than the 
width of a SM Higgs of similar mass \cite{LittRad}.
For other work on warped models with a decoupled KK sector ($\mkk\gg 1$~TeV)  see
Refs.~\cite{WTCM,WTCMrad}.

\section{Setup}  

We will adopt the usual RS background metric \cite{Randall:1999ee}
\beq
ds^2 = e^{-2k y} \eta_{\mu \nu}d x^\mu d x^\nu - dy^2\,, 
\label{RSmetric}
\eeq
where $k$ is the curvature scale, typically
assumed smaller than the 5D fundamental  scale $M_5$.  The compact
dimension $y$ is bounded by ultraviolet (UV) and infrared (IR)
branes at $y=0,L$, respectively.  The gauge and fermion content of
the SM are placed in the 5D bulk.  We will not require any other
gauge symmetries beyond the SM $SU(3)_C \times SU(2)_L \times
U(1)_Y$. The electroweak symmetry is assumed to be broken by an
IR-brane-localized Higgs doublet\footnote{We note that the bulk
Higgs in warped gauge-Higgs unification models \cite{comph1,comph2}
receives 1-loop mass corrections cut off by KK masses and is less
fine-tuned.  However, these models are in general subject to the
same severe tensions with the flavor data that push the KK scale to
$\sim 10$ TeV.}. The flavor structure of the SM can be obtained,
using bulk fermions with non-zero vector-like masses $m_i$, $i=u, d,
\ldots$ \cite{Grossman:1999ra,Gherghetta:2000qt}. The resulting
zero-mode fermions are exponentially localized in 5D, parameterized
by $c_i\equiv m_i/k$, with $c_i\sim 1$ for light fermions that are
UV-localized and have small overlaps with the IR-localized Higgs.

The radion $\phi$ represents \cite{GW1} quantum fluctuations of the
position of the IR brane and interacts through its couplings to the
trace of the energy-momentum tensor; these couplings are suppressed
by the scale \cite{GW2}
\beq
\Lphi \equiv e^{-k L} \sqrt{6 M_5^3/k}.  
\label{Lphi}\eeq
The interactions of $\phi$ with
bulk fields are derived in Ref.~\cite{CHL} and summarized in
Ref.~\cite{LittRad}, to which we refer the interested reader for the
relevant expressions and details \cite{Early-radion}.  In this work, for simplicity, we
will not consider possible brane-localized kinetic terms, as their
inclusion will not change our results qualitatively.  
We will also ignore Higgs-radion mixing (for an early discussion of this possibility see 
the third work in Ref.~\cite{Early-radion}).  This mixing is
proportional to $\vev{H}/\Lambda_\phi$ and for 
$\Lambda_\phi \sim 10$~TeV (as we have typically assumed in our work) 
it is a very small effect and can be ignored in our study.  
We note that interactions of the radion that are relevant to our analysis
are governed by the low-energy states in the theory.  Hence, the
details of bulk gauge symmetries are not very important here, and our
assumption of a SM bulk gauge content leads to conclusions that apply
also to other more complicated scenarios.  
For some recent works on radion phenomenology 
see, for example, Ref.~\cite{Recent-radion}.

As a guide for phenomenology, we will
consider the Goldberger-Wise (GW) mechanism
\cite{GW1,GW2}, with a bulk scalar $\Phi$ of mass $m$ and brane-localized
potentials.  The 5D vacuum expectation
values of $\Phi$ on the UV and the IR branes are denoted by  $v_0$ and
$v_L$ (with mass dimension 3/2), respectively.
The stabilized radius $L$ is then given by \cite{GW1,GW2}
\beq
k L = \epsilon^{-1} \ln(v_0/v_L),
\label{GWkL}
\eeq
where
$\epsilon\equiv m^2/(4 k^2)$ and
\beq
m_\phi^2 =
\frac{v_L^2}{3 M_5^3}\, \epsilon^2 \tilde{k}^2\,,
\label{GWmphi}
\eeq
with $\ktil \equiv k e^{-kL}$ the warped-down curvature scale.

\section{Electroweak Constraints} 

Various corrections resulting from the appearance of new states above the weak scale
can be parametrized in terms of the oblique Peskin-Takeuchi
$(S,T)$ parameters \cite{Peskin:1991sw} and we will discuss
them below.  Contributions from the tree-level mixing of
the gauge zero modes with the heavy KK modes
are given by \cite{custodial1},
\beq
S_{\rm tree} \approx 2\pi\, (\vev{H}/\ktil)^2\left[1 - \frac{1}{k L} + \xi(c)\right]
\label{St}
\eeq
and
\beq
T _{\rm tree} \approx \frac{\pi}{2 \cos\theta_W^2}(\vev{H}/\ktil)^2
\left[k L - \frac{1}{k L} + \xi(c)\right],
\label{Tt}
\eeq
where
\beq
\xi(c)\equiv
\frac{(2c - 1)/(3 - 2c)}{1 - e^{k L (2c-1)}}
\left(2k L - \frac{5-2c}{3-2c}\right)
\label{xi}
\eeq
is a function of fermion localization parameter $c$ and 
$\cos^2\theta_W\simeq 0.77$.  For fermion profiles that lead to a realistic flavor  pattern we have $\xi(c) \ll 1$.

In the absence of a 5D custodial symmetry, a UV-sensitive
loop contribution to the $T$ parameter arises.  This
dependence on the cutoff-scale physics can be ``renormalized'' by the addition of
a higher-dimension operator.  One can use na\"ive dimensional analysis relevant for
strong dynamics \cite{Georgi:1992dw} to estimate the size of the UV-sensitive contribution by
\beq
O_{\rm UV} \sim \frac{(D^\mu H)^\dagger H(H^\dagger D_\mu H)}{\ktil^2},
\label{OUV}
\eeq
where $\ktil$ plays the role of the decay constant for a composite particle \cite{comph2}.
The contribution from the above operator to the $T$ parameter can then be estimated
by \cite{custodial1}
\beq
T_{\rm UV} \sim \frac{1}{2 \alpha}\left(\frac{\vev{H}}{\ktil}\right)^2\,,
\label{TUV}
\eeq
where $\alpha$ is the electromagnetic fine-structure constant.  
The results from Refs.~\cite{Carena:2006bn,Carena:2007ua}  
suggest that the loop contributions to the $S$ parameter 
summed over the KK modes are not large, even for $\sim 3$~TeV KK
masses they consider.

\section{Range of parameters} 

We will assume that the Higgs is
light: $m_H \sim 125$~GeV (other values in this range will also lead to
nearly the same conclusions reached below).  We now examine the expected sizes
of $\delta T$ and $\delta S$ in the deca-TeV warped model considered
in this work.  For the sake of concreteness, let us consider $\mkk = 10$~TeV
for bulk gauge fields, which implies $\ktil \simeq 4$~TeV
\cite{Davoudiasl:1999tf,Pomarol:1999ad}.  The value of $kL$
determines the UV scale $k$ in the RS geometry through $k = \ktil
e^{kL}$.  We will consider a range of values bounded by 
$kL=10$ and $kL=30$.  With $kL=10$, we have $k\sim 10^5$~TeV, corresponding 
to a ``Little'' RS scenario for flavor \cite{LRS}; note that this value for $k$ is sufficiently large that the resulting 
model can avoid conflict with even the most stringent flavor constraints \cite{Bauer:2008xb}.  For $kL=30$, we get 
$k\sim 10^{16}$~GeV, close to $\mP$ and similar to the original setup \cite{Randall:1999ee}. 

For the above choice of parameters, Eq.~(\ref{St}) then implies $\delta S \simeq
0.02$, and for $\delta T = T_{\rm tree} + T_{\rm UV}$, we find
$\delta T \sim 0.3\text{--}0.5$ for $10 \leq kL \leq 30$.  Hence,
agreement with electroweak data \cite{GfitterOb} may require a bulk custodial
symmetry or a somewhat larger KK scale.  Alternatively, the Higgs could
be heavy, say above $\sim 600$~GeV \cite{GfitterOb,Peskin:2001rw}; one may consider
this possibility if the present hints for a light Higgs do not
persist with more data
\cite{note-added}.  
In any event, our main result---that a sole weak-scale radion (dilaton) 
can provide indirect evidence for KK (composite) states at scales as high 
as $\sim 10$~TeV---does not depend sensitively on the mass of the Higgs.  

For simplicity, we will set $k=M_5$, which gives $\Lphi=\sqrt{6}\,\ktil$; hence we will have $\Lphi\simeq\mkk$.   
Our choice for $k$ is consistent with ignoring higher-order terms    
in 5D curvature $|R_5| = 20 k^2$ \cite{Davoudiasl:2000wi}, as assumed in derivation of the RS background 
\cite{Randall:1999ee}, where the expansion parameter is $R_5/M_c^2$ and $M_c \sim \sqrt[3]{24}\,\pi M_5$ \cite{M5}. 
Since $v_0$ and $v_L$ are 5D parameters, it is reasonable to assume that
$v_{0,L}\sim k^{3/2}$ and $ \ln(v_0/v_L) \sim 1$, which implies $\epsilon
\sim (k L)^{-1}$, from Eq.~(\ref{GWkL}). Using Eq.~(\ref{GWmphi}), one then
finds $m_\phi \sim \tilde{k}/(k L)$.  Hence, for $10 \leq kL \leq 30$ we may expect 
$m_\phi$ to be of order a few hundred~GeV in our setup.
\footnote{If the IR brane tension is ``detuned'' significantly from the RS background value, the radion mass scaling can be changed to $m_\phi\sim\tilde k/\sqrt{kL}$ \cite{Konstandin:2010cd,Eshel:2011wz}, in which case the radion could be somewhat heavier: $m_\phi\sim500\text{--}1000~\text{GeV}$.
The typical radion masses considered in our analysis may then require that the IR brane tension is somewhat tuned.
In any event, these simple estimates ignore order-unity factors  
coming, for example, from the specific parameters of the stabilizing scalar potential.  
Hence, the mass range in our analysis may be relevant even in the case of large IR brane tension detuning, 
but this depends on the specifics of the stabilization mechanism that lie 
outside the scope of our phenomenological analysis.  
We thank K.~Agashe for emphasizing these issues.}

\section{Results}

The radion can be singly produced at the LHC via gluon fusion: $gg\to\phi$.
The partonic cross section is given by
\begin{equation}
\hat\sigma(gg\to\phi)=\frac\pi4C_{gg}^2\frac{m_\phi^2}{\Lambda^2}\delta(\hat s-m_\phi^2),
\end{equation}
where $\sqrt{\hat s}$ is the partonic center-of-mass energy and $C_{gg}=-1/(4kL)-23\alpha_s/(24\pi)$ if $m_\phi<2m_t$.
This may be compared to the cross section for production of a SM Higgs boson in the $m_t\to\infty$ limit \cite{Georgi:1977gs}:
\begin{equation}
\hat\sigma(gg\to H)=\frac{\alpha_s^2}{576\pi}\frac{m_H^2}{v^2}\delta(\hat s-m_H^2),
\end{equation}
where $v$ is the vacuum expectation value of the Higgs field.  Hence, in the regime of 
validity of the above equations, we have 
\begin{equation}
\hat\sigma(gg\to\phi)=\left (\frac{12 \pi \, C_{gg}\, v }{\alpha_s \,\Lambda}\right )^2 \hat\sigma(gg\to H).
\label{ggphih}
\end{equation}
The above equation suggests that $\hat\sigma(gg\to\phi) \sim 0.1\, \hat\sigma(gg\to H)$ for $kL=30$ and $\Lphi=10~\text{TeV}$.
The Higgs production cross section via gluon fusion at the 14 TeV LHC for $m_H=125$~GeV, 
for example, is about 50~pb \cite{HiggsCSWG}, which includes a $K$-factor of $\sim 2$ 
from next-to-next-to-leading order \cite{Harlander:2002wh} and next-to-next-to-leading 
logarithm \cite{Catani:2003zt} corrections.  We find 
that the corresponding leading order 
cross section for $m_\phi=125$~GeV is about 1.8~pb, which is 
consistent with the na\"ive expectation from Eq.~(\ref{ggphih}).  

\begin{figure}[bt]
\includegraphics[width=0.7\textwidth]{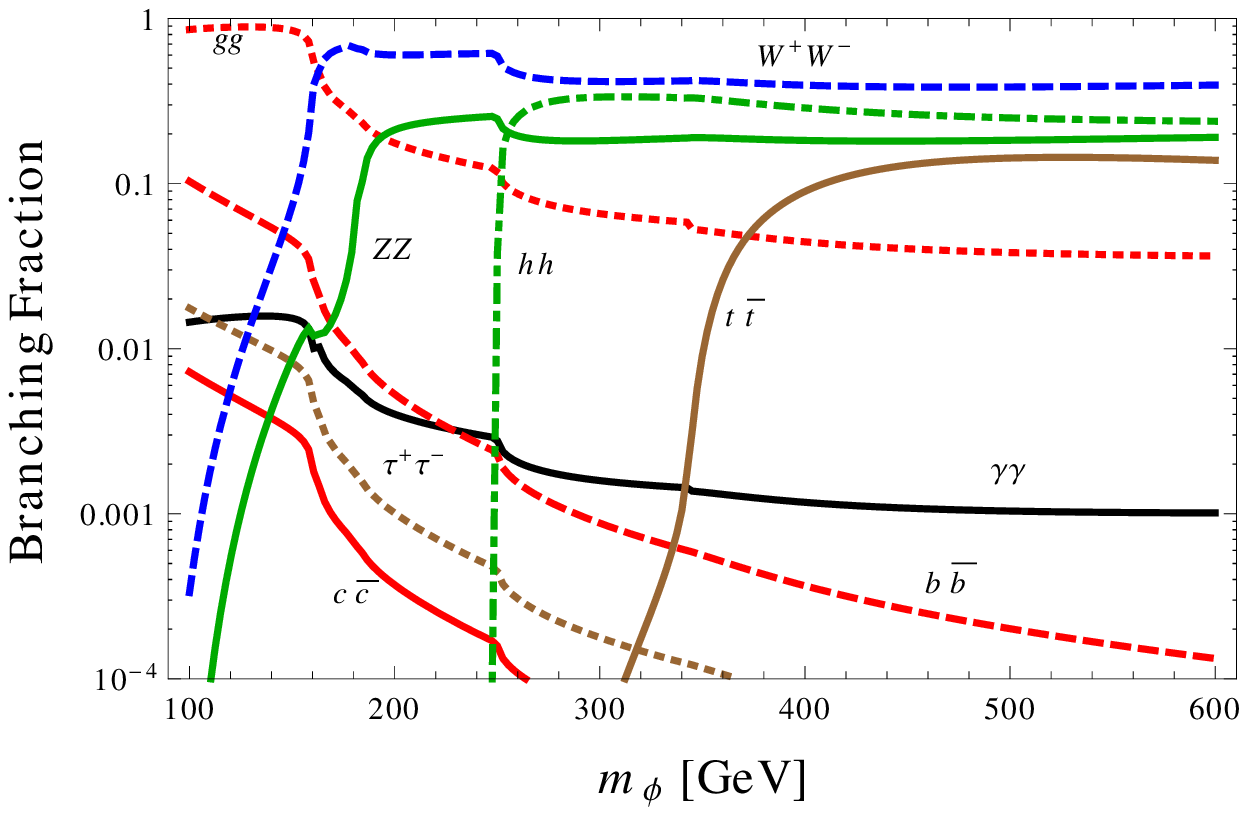}
\caption{Branching fractions of the radion as a function of radion mass, assuming $m_H=125~\text{GeV}$, $kL=10$, and $\Lambda_\phi=10~\text{TeV}$.}
\label{f:branch}
\end{figure}

Provided the radion is sufficiently heavy ($m_\phi\gsim2m_W$), its dominant decay mode is to a pair of $W$ bosons.
See, for example, Fig.~\ref{f:branch}, illustrating the branching fractions of the radion for one choice of parameters.

We first consider a search for the radion in the 
$WW$ channel at the LHC, following the planned energy upgrade to $14~\text{TeV}$.
In order to minimize QCD background, we take as our signal process the fully leptonic channel: $gg\to\phi\to W^+W^-\to l^+\nu_ll'^-\bar\nu_{l'}$, where $l$ and $l'$ may be either $e$ or $\mu$.
We compute this process at leading order in the narrow-width approximation, using the \code{Cuba} library \cite{Hahn:2004fe} for numerical integration.
The irreducible background is the SM process $pp\to l^+l'^-\nu\bar\nu'$ (dominated by SM $WW$ production), which we simulate using \code{MadGraph 5} \cite{Alwall:2011uj}.  
Both signal and background are computed using the CT10 parton distributions \cite{Lai:2010vv}.

We impose the following cuts, somewhat similar to those used in Higgs searches at the LHC \cite{CMS-HWW,ATLAS-HWW}.
We require exactly two oppositely charged leptons ($e$ or $\mu$), each with pseudorapidity $|\eta|<2.5$, and no accompanying jets.
One of the leptons must have transverse momentum $p_T>20~\text{GeV}$, while the other must have $p_T>15~\text{GeV}$.
The two leptons must have an invariant mass $\mll>10~\text{GeV}$ and be separated by $\Delta R>0.4$, where $\Delta R\equiv\sqrt{(\Delta\varphi)^2+(\Delta\eta)^2}$ is the separation in azimuthal angle $\varphi$ and pseudorapidity $\eta$.
When both leptons have the same flavor ($e^+e^-$ or $\mu^+\mu^-$), we further require that $\mll>15~\text{GeV}$ and $|\mll-m_Z|>15~\text{GeV}$, in order to suppress the Drell-Yan background.
Additionally, we require large missing transverse energy $\ETmiss$, which we identify as the vector sum of the neutrinos' transverse momenta: $\ETmiss>25~\text{GeV}$ for $e^\pm\mu^\mp$ events and $\ETmiss>45~\text{GeV}$ for $e^+e^-$ and $\mu^+\mu^-$ events.

\newcommand\reachcaption[1]{The $3\sigma$ (dashed) and $5\sigma$ (solid) contours, in the #1 plane, for $\phi\to W^+W^-\to l^+l^-\nu\bar\nu$ at the LHC with $100~\invfb$ at $14~\text{TeV}$}

\begin{figure}[ptb]
\includegraphics[width=0.5\textwidth]{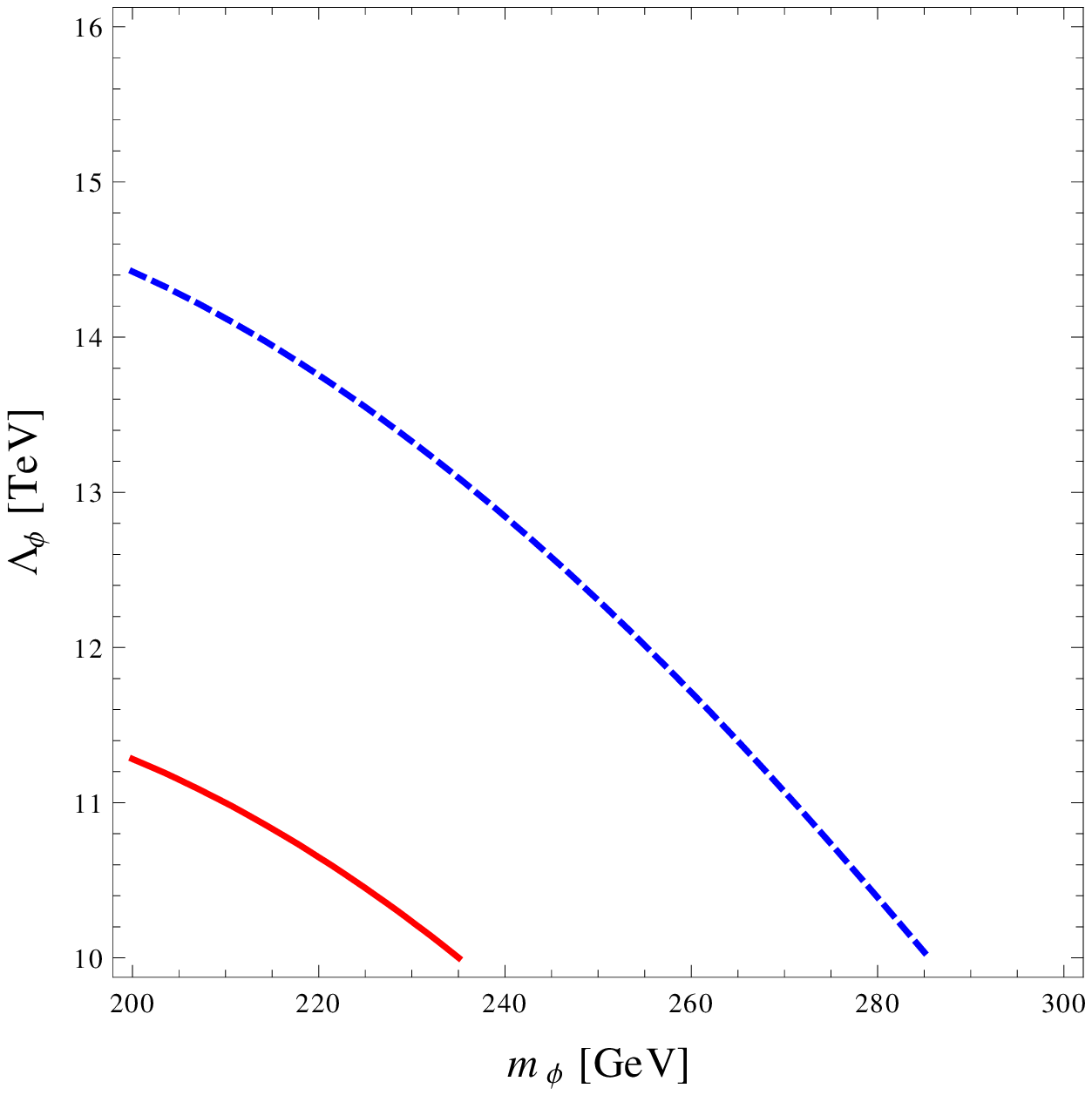}
\caption{\reachcaption{$(m_\phi,\Lphi)$}, with $kL=10$.}
\label{f:reach_m_Lam}
\end{figure}

\begin{figure}[ptb]
\includegraphics[width=0.5\textwidth]{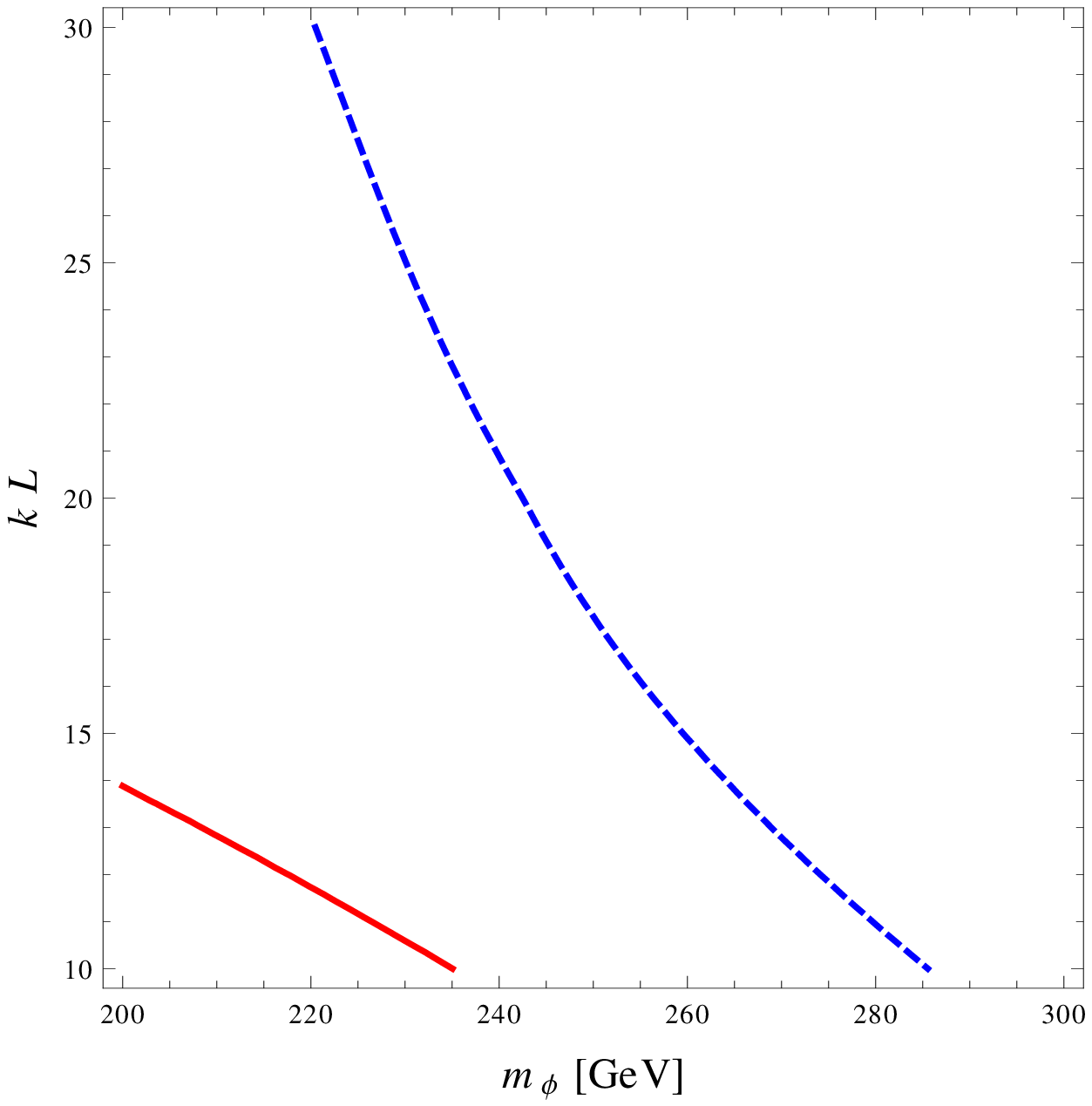}
\caption{\reachcaption{$(m_\phi,kL)$}, with $\Lphi=10~\text{TeV}$.}
\label{f:reach_m_kL}
\end{figure}

\begin{figure}[ptb]
\includegraphics[width=0.5\textwidth]{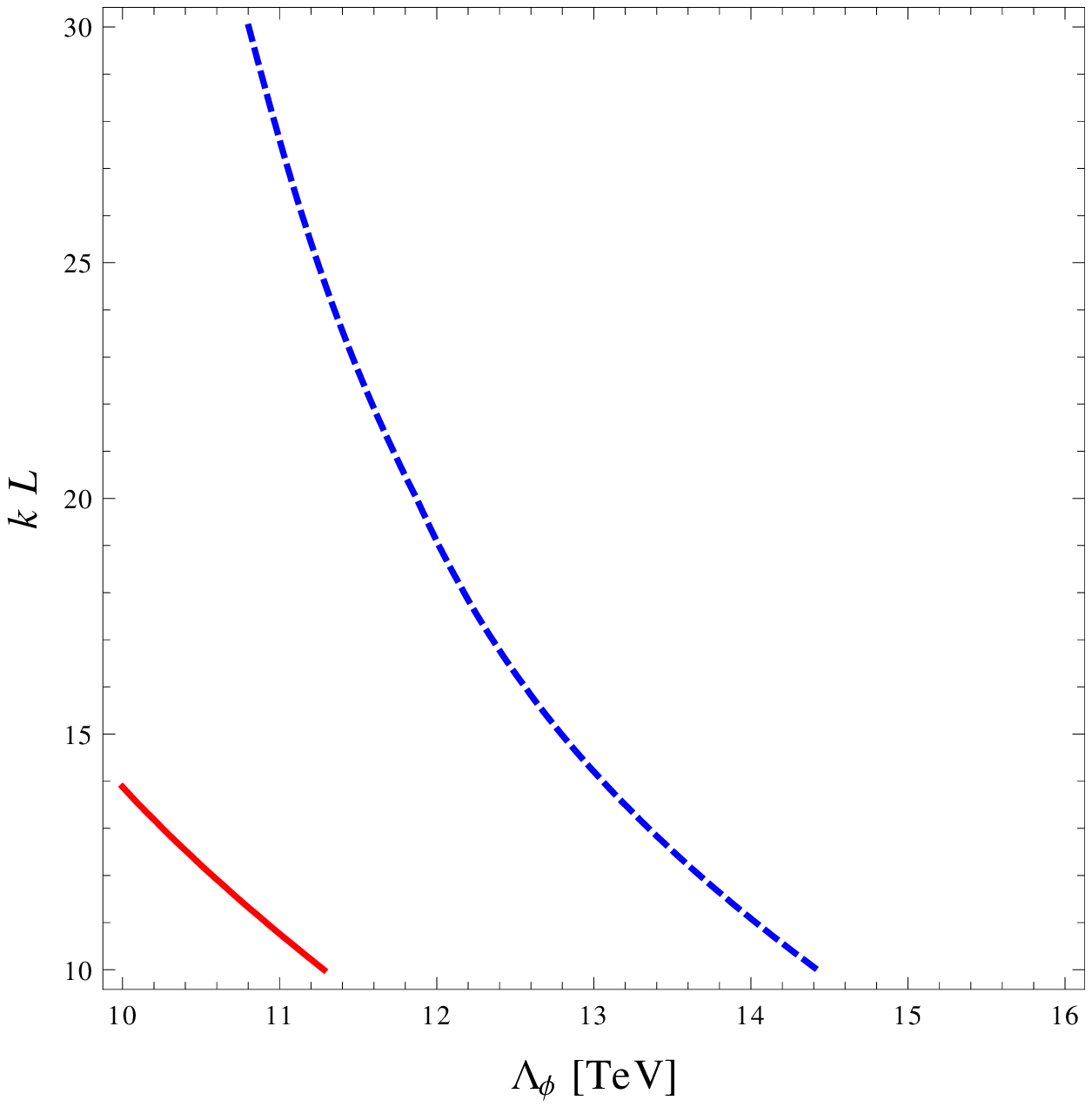}
\caption{\reachcaption{$(\Lphi,kL)$}, with $m_\phi=200~\text{GeV}$.}
\label{f:reach_Lam_kL}
\end{figure}

Finally, we consider a transverse mass variable $m_T$, defined by
\begin{equation}\label{eq:mT}
m_T^2\equiv\left(\sqrt{|\pTll|^2+\mll^2}+\ETmiss\right)^2-|\pTll+\pTmiss|^2,
\end{equation}
where $\pTll$ is the transverse momentum of the lepton pair, $\pTmiss$ is the missing transverse momentum, and $\ETmiss=|\pTmiss|$ \cite{ATLAS-HWW,Barr:2009mx}.
The definition of $m_T$ is such that $m_T\leq m_\phi$ for all signal events.
Because of this relation between $m_T$ and $m_\phi$, the distribution of $m_T$ can be used to provide an estimate of $m_\phi$.
It may be possible to obtain an improved estimate by considering alternative transverse-mass variables that bound $m_\phi$ more tightly \cite{Barr:2011si}.
However, in this work we restrict our attention to $m_T$ as defined in Eq.~\eqref{eq:mT};
in order to test for the presence of a radion with mass $m_\phi$, we require that $m_\phi/2<m_T<m_\phi$. 

The model parameters relevant for this search are $m_\phi$, $\Lphi$, and $kL$.
In Figs.~\ref{f:reach_m_Lam}--\ref{f:reach_Lam_kL}, we show $3\sigma$ and $5\sigma$ contours in various slices of this parameter space, assuming $100~\invfb$ of integrated luminosity at the LHC with a center-of-mass energy of $14~\text{TeV}$.
The significance is defined as $S/\sqrt B$, where $S$ and $B$ respectively denote the numbers of signal and background events surviving the cuts.
The expected numbers of signal and background events are shown, for a few representative points in parameter space, in Table~\ref{t:events}.

\begin{table}
\begin{tabular}{ccc|cc|c}
$m_\phi/\text{GeV}$ & $\Lphi/\text{TeV}$ & $kL$ & Signal & Background & $S/\sqrt B$ \\
\hline
200 & 10 & 10 & $1.57\times10^3$ & $6.49\times10^4$ & 6.18 \\
300 & 10 & 10 & 557 & $4.81\times10^4$ & 2.54 \\
200 & 15 & 10 & 700 & $6.49\times10^4$ & 2.75 \\
200 & 10 & 30 & 873 & $6.49\times10^4$ & 3.43
\end{tabular}
\caption{The expected numbers of signal and background events passing the cuts, and the significance $S/\sqrt B$, for selected values of model parameters, at the LHC with $100~\invfb$ at $14~\text{TeV}$.}\label{t:events}
\end{table}

\begin{figure}[phtb]
\includegraphics[width=0.5\textwidth]{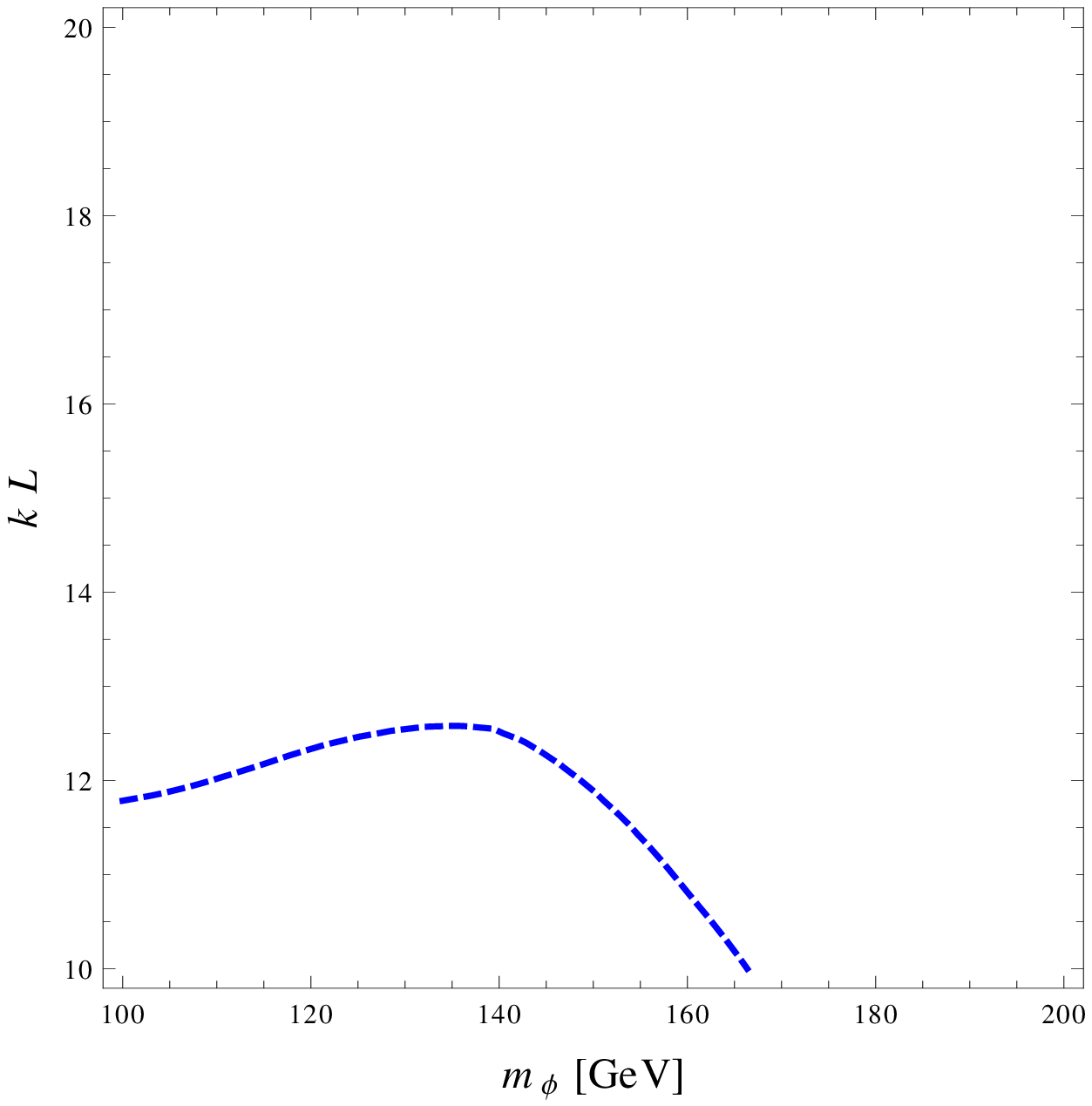}
\caption{The $3\sigma$ contour, in the $(m_\phi,kL)$ plane, for $\phi\to\gamma\gamma$ at the LHC with $100~\invfb$ at $14~\text{TeV}$, with $\Lphi=10~\text{TeV}$.}
\label{f:diphoton}
\end{figure}

For radion masses below the $WW$ threshold, an important search channel is the diphoton final state, especially 
for smaller values of $kL$ \cite{LittRad} \footnote{We note that, for values of $kL$ 
in the lower part of the range considered here, the branching fraction for $\phi \to \gamma\gamma$ tends to be 
significantly larger than the corresponding branching fraction of the SM Higgs; see Fig.~\ref{f:branch}.}.  
The observation of the radion signal in this channel would provide the value 
of $m_\phi$ through the reconstruction of the resonant peak.  
In Fig.~\ref{f:diphoton}, assuming $\Lambda_\phi = 10$~TeV, we have plotted  the $3\sigma$ reach for this channel in the $(m_\phi, kL)$ plane, using the methodology of Ref.~\cite{LittRad} 
and assuming $100~\invfb$ of integrated luminosity at $14~\text{TeV}$.
We see that for $kL\lsim12$, significant evidence for a radion of mass 
in the range 100--160~GeV can be obtained.  Therefore we find that, through the $\gamma \gamma$ and $WW$ channels, 
the LHC has the potential to detect a radion signal over a healthy portion of parameter space, probing radion masses up to $m_\phi\sim290~\text{GeV}$ and scales as high as $\Lphi\sim14~\text{TeV}$.

In case the current hints for a Higgs at about 125~GeV persist with more data, we see 
from Fig.~\ref{f:branch} that $\phi \to h h$ is one of the dominant 
decay channels of the radion for $m_\phi \gsim 250$~GeV.  
If the Higgs is sufficiently SM-like, we may expect that each Higgs will mainly decay into 
a $b\bar b$ pair.  This signal suffers from a large $4 b$ jet QCD background \cite{4bjets}.  
While one may devise suitable cuts in order to make the $4 b$ final state a useful 
search channel \cite{Lafaye:2000ec},  looking for the radion using 
this final state will likely require improved analysis techniques and a detailed study, 
which lie outside the scope of this paper.

We close this section with a comment on the possibility of identifying the radion.  If a narrow 
scalar is found at a few hundred GeV, 
in principle, measurements of its branching fractions could be a guide to its identity.  For example, in the context of 
RS-like models of flavor, as examined here, we may expect a typical set of branching fractions comparable to those 
presented in our Fig.~\ref{f:branch}.  However, it should be kept in mind that due to various model-dependent assumptions, 
one cannot make very precise statements here.  What we have tried to demonstrate in our work is that, even if the 
scale of the new physics is at about 10 TeV, one may still have access to the radion 
signal and a hint for a nearby scale, in the class of models we have considered.

\section{Conclusions}
 
In this work, we considered the possibility that the threshold for new phenomena may be at the deca-TeV (10 TeV) scale, as 
suggested by indirect precision measurements.  In such a circumstance, 
one may ask whether there are physics scenarios that are 
governed  by scales as high as 10 TeV, but also include light signature states that are accessible at the LHC energies.  
Good examples of such scenarios are the 5D warped models of flavor based on the original Randall-Sundrum (RS) background.  
The simplest versions of such 
models give rise to KK states whose masses are naturally pushed to scales of order 10 TeV, if they are to avoid 
disagreement with precision electroweak and flavor data.  In order to lower the KK masses to a few TeV, these models must be augmented by a number of new gauge symmetries and large additions to their field content, leading to quite complicated setups.  Discovery of a SM-like Higgs, 
hints for which may have been detected in the 2011 LHC data,  
will strengthen the case for a roughly 10~TeV lower bound on the KK threshold.

While the LHC will not have direct access to the deca-TeV KK states, we showed in this work that the radion scalar, associated 
with the quantum fluctuation of the compact fifth dimension, could very well be discovered at the LHC, 
with design parameters.  We considered realistic warped flavor scenarios, characterized by UV 
scales $\sim 10^5\text{--}10^{13}~\text{TeV}$ and KK masses 
of $\sim 10\text{--}15~\text{TeV}$.  We focused on the gluon-fusion production of the  
radion.   For $m_\phi > 2 m_W$, we considered the typically 
dominant $WW$ decay channel, followed by leptonic decays of each $W$.  
For $m_\phi < 2 m_W$, we examined the utility of the diphoton 
channel in searching for the radion.  Our analysis indicates that 
a radion of mass $\sim 100\text{--}300~\text{GeV}$ can be detected by the LHC experiments at the $\sim (3\text{--}5)\sigma$ level, for interesting parameter ranges of warped flavor models, assuming 14~TeV 
for the center-of-mass energy and $\sim 100$~\invfb\ of data.
Other decay channels, such as $\phi\to WW\to l\nu jj$ and $\phi\to ZZ$, can be included in a more comprehensive analysis, leading to an improved reach.  However, our results give a good estimate of the possibilities at the LHC.  
We also pointed out that assuming 
a SM-like Higgs at $\sim 125$~GeV, one may consider the $\phi\to h h \to b\bar b b \bar b$ signal for 
$m_\phi \gsim 250$~GeV, but this will likely require improvements in analysis 
techniques to control the large QCD background.

Our conclusions suggest that, through the production of a weak scale radion, experimental evidence for a warped deca-TeV threshold could be accessible at the LHC in coming years.  
Similar statements are applicable to dual 4D theories, with a dynamical 
scale around 10~TeV, whose spectrum is expected to include a light dilaton associated with spontaneous 
conformal symmetry breaking.  In either picture, the discovery of a light and narrow scalar can herald the 
appearance of new physics at the deca-TeV scale, motivating new experiments at center-of-mass energies beyond 
that of the LHC.

{\it Note added

After this work was completed and during the review process, ATLAS \cite{:2012gk} and CMS \cite{:2012gu} 
announced the discovery of a Higgs-like 
state at about 125~GeV.  More data is required to determine, at a statistically significant level, whether this new state 
has properties that are different from those of the SM Higgs.  However, the possibility of a heavy Higgs 
above $\sim$~600~GeV, mentioned earlier in our discussion, is now strongly disfavored.

}


\acknowledgments

We thank K.~Agashe, S.~Dawson, and E.~Pont\'on for discussions.
The work of H.D.~and A.S.~is supported in part by the US DOE Grant DE-AC02-98CH10886.
The work of T.M.~is supported by the Department of Energy under Award Number DE-FG02-91ER40685.



\begin{thebibliography}{99}

\bibitem{LEPEWWG}
LEP Electroweak Working Group,{\tt http://lepewwg.web.cern.ch/LEPEWWG/\,.}

\bibitem{GfitterSM}
GFitter Group, {\tt http://gfitter.desy.de/Standard\_Model/\,.}

\bibitem{LEP-Higgs}
  R.~Barate {\it et al.} [ LEP Working Group for Higgs boson searches and ALEPH and DELPHI and L3 and OPAL Collaborations ],
  Phys.\ Lett.\  {\bf B565}, 61-75 (2003)
  [hep-ex/0306033].

\bibitem{uh1}
  B.~W.~Lee, C.~Quigg, H.~B.~Thacker,
  Phys.\ Rev.\ Lett.\  {\bf 38}, 883-885 (1977).

\bibitem{uh2}
  W.~J.~Marciano, G.~Valencia, S.~Willenbrock,
  Phys.\ Rev.\  {\bf D40}, 1725 (1989).

\bibitem{ATLAS-Higgs}
  G.~Aad {\it et al.}  [ATLAS Collaboration],
  Phys.\ Lett.\ B {\bf 710}, 49 (2012)
  [arXiv:1202.1408 [hep-ex]].

\bibitem{CMS-Higgs}
  S.~Chatrchyan {\it et al.}  [CMS Collaboration],
  Phys.\ Lett.\ B {\bf 710}, 26 (2012)
  [arXiv:1202.1488 [hep-ex]].

\bibitem{note-added}
See the note added at the end of this paper.

\bibitem{Randall:1999ee}
  L.~Randall and R.~Sundrum,
  Phys.\ Rev.\ Lett.\  {\bf 83}, 3370 (1999)
  [arXiv:hep-ph/9905221].

\bibitem{Davoudiasl:1999tf}
  H.~Davoudiasl, J.~L.~Hewett and T.~G.~Rizzo,
  Phys.\ Lett.\  B {\bf 473}, 43 (2000)
  [arXiv:hep-ph/9911262];

\bibitem{Pomarol:1999ad}
  A.~Pomarol,
  Phys.\ Lett.\  B {\bf 486}, 153 (2000)
  [arXiv:hep-ph/9911294].

\bibitem{Grossman:1999ra}
  Y.~Grossman and M.~Neubert,
  Phys.\ Lett.\  B {\bf 474}, 361 (2000)
  [arXiv:hep-ph/9912408].

\bibitem{Gherghetta:2000qt}
  T.~Gherghetta and A.~Pomarol,
  Nucl.\ Phys.\  B {\bf 586} (2000) 141
  [arXiv:hep-ph/0003129].

\bibitem{Carena:2003fx} 
  M.~S.~Carena, A.~Delgado, E.~Pont\'on, T.~M.~P.~Tait and C.~E.~M.~Wagner,
  Phys.\ Rev.\ D {\bf 68}, 035010 (2003)
  [hep-ph/0305188].

\bibitem{custodial1}
  K.~Agashe, A.~Delgado, M.~J.~May and R.~Sundrum,
  JHEP {\bf 0308}, 050 (2003)
  [arXiv:hep-ph/0308036].

\bibitem{custodial2}
  K.~Agashe, R.~Contino, L.~Da Rold and A.~Pomarol,
  Phys.\ Lett.\  B {\bf 641}, 62 (2006)
  [arXiv:hep-ph/0605341];

\bibitem{Carena:2006bn} 
  M.~S.~Carena, E.~Ponton, J.~Santiago and C.~E.~M.~Wagner,
  Nucl.\ Phys.\ B {\bf 759}, 202 (2006)
  [hep-ph/0607106].

\bibitem{Carena:2007ua}
  M.~S.~Carena, E.~Pont\'on, J.~Santiago and C.~E.~M.~Wagner,
  Phys.\ Rev.\  D {\bf 76}, 035006 (2007)
  [arXiv:hep-ph/0701055].

\bibitem{Bona:2007vi}
  M.~Bona {\it et al.} [ UTfit Collaboration ],
  JHEP {\bf 0803}, 049 (2008).
  [arXiv:0707.0636 [hep-ph]].


\bibitem{fb1}
  C.~Cs\'aki, A.~Falkowski and A.~Weiler,
  JHEP {\bf 0809}, 008 (2008)
  [arXiv:0804.1954 [hep-ph]].

\bibitem{fb2}
  O.~Gedalia, G.~Isidori and G.~Perez,
  Phys.\ Lett.\  B {\bf 682}, 200 (2009)
  [arXiv:0905.3264 [hep-ph]].

\bibitem{fb3} 
  S.~Casagrande, F.~Goertz, U.~Haisch, M.~Neubert and T.~Pfoh,
  JHEP {\bf 0810}, 094 (2008)
  [arXiv:0807.4937 [hep-ph]]; 
  M.~Bauer, S.~Casagrande, U.~Haisch and M.~Neubert,
  JHEP {\bf 1009}, 017 (2010)
  [arXiv:0912.1625 [hep-ph]].


\bibitem{Blanke:2008zb} 
  M.~Blanke, A.~J.~Buras, B.~Duling, S.~Gori and A.~Weiler,
  JHEP {\bf 0903}, 001 (2009)
  [arXiv:0809.1073 [hep-ph]].

\bibitem{Bauer:2011ah}
  M.~Bauer, R.~Malm and M.~Neubert,
  Phys.\ Rev.\ Lett.\  {\bf 108}, 081603 (2012)
  [arXiv:1110.0471 [hep-ph]].

\bibitem{Azatov:2010pf}
  A.~Azatov, M.~Toharia and L.~Zhu,
  Phys.\ Rev.\ D {\bf 82}, 056004 (2010)
  [arXiv:1006.5939 [hep-ph]].

\bibitem{Goertz:2011hj} 
  F.~Goertz, U.~Haisch and M.~Neubert,
  Phys.\ Lett.\ B {\bf 713}, 23 (2012)
  [arXiv:1112.5099 [hep-ph]].

\bibitem{Carena:2012fk} 
  M.~Carena, S.~Casagrande, F.~Goertz, U.~Haisch and M.~Neubert,
  JHEP {\bf 1208}, 156 (2012)
  [arXiv:1204.0008 [hep-ph]].

\bibitem{LHCKK}
See, for example: 
  K.~Agashe, A.~Belyaev, T.~Krupovnickas, G.~Perez and J.~Virzi,
  Phys.\ Rev.\ D {\bf 77}, 015003 (2008)
  [hep-ph/0612015]; 
A.~L.~Fitzpatrick, J.~Kaplan, L.~Randall and L.~-T.~Wang,
  JHEP {\bf 0709}, 013 (2007)
  [hep-ph/0701150]; 
B.~Lillie, L.~Randall and L.~-T.~Wang,
  JHEP {\bf 0709}, 074 (2007)
  [hep-ph/0701166]; 
K.~Agashe, H.~Davoudiasl, G.~Perez and A.~Soni,
  Phys.\ Rev.\ D {\bf 76}, 036006 (2007)
  [hep-ph/0701186]; 
A.~Djouadi, G.~Moreau and R.~K.~Singh,
  Nucl.\ Phys.\ B {\bf 797}, 1 (2008)
  [arXiv:0706.4191 [hep-ph]]; 
K.~Agashe, H.~Davoudiasl, S.~Gopalakrishna, T.~Han, G.-Y.~Huang, G.~Perez, Z.-G.~Si and A.~Soni,
  Phys.\ Rev.\ D {\bf 76}, 115015 (2007)
  [arXiv:0709.0007 [hep-ph]]; 
O.~Antipin, D.~Atwood and A.~Soni,
  Phys.\ Lett.\ B {\bf 666}, 155 (2008)
  [arXiv:0711.3175 [hep-ph]]; 
K.~Agashe, S.~Gopalakrishna, T.~Han, G.-Y.~Huang and A.~Soni,
  Phys.\ Rev.\ D {\bf 80}, 075007 (2009)
  [arXiv:0810.1497 [hep-ph]]; 
H.~Davoudiasl, T.~G.~Rizzo and A.~Soni,
  Phys.\ Rev.\ D {\bf 77}, 036001 (2008)
  [arXiv:0710.2078 [hep-ph]]; 
H.~Davoudiasl, S.~Gopalakrishna and A.~Soni,
  Phys.\ Lett.\ B {\bf 686}, 239 (2010)
  [arXiv:0908.1131 [hep-ph]].



\bibitem{EarlyHiggs}
See, also: 
B.~Lillie,
  JHEP {\bf 0602}, 019 (2006)
  [hep-ph/0505074];
A.~Djouadi and G.~Moreau,
  Phys.\ Lett.\ B {\bf 660}, 67 (2008)
  [arXiv:0707.3800 [hep-ph]];
S.~Casagrande, F.~Goertz, U.~Haisch, M.~Neubert and T.~Pfoh,
  JHEP {\bf 1009}, 014 (2010)
  [arXiv:1005.4315 [hep-ph]];
C.~Bouchart and G.~Moreau,
  Phys.\ Rev.\ D {\bf 80}, 095022 (2009)
  [arXiv:0909.4812 [hep-ph]];
G.~Cacciapaglia, A.~Deandrea and J.~Llodra-Perez,
  JHEP {\bf 0906}, 054 (2009)
  [arXiv:0901.0927 [hep-ph]].

\bibitem{Agashe:2004ay}
  K.~Agashe, G.~Perez, A.~Soni,
  Phys.\ Rev.\ Lett.\  {\bf 93}, 201804 (2004)
  [hep-ph/0406101].

\bibitem{Agashe:2004cp}
  K.~Agashe, G.~Perez, A.~Soni,
  Phys.\ Rev.\  {\bf D71}, 016002 (2005)
  [hep-ph/0408134].

\bibitem{ADSCFT}
  J.~M.~Maldacena,
  Adv.\ Theor.\ Math.\ Phys.\  {\bf 2}, 231 (1998)
  [Int.\ J.\ Theor.\ Phys.\  {\bf 38}, 1113 (1999)]
  [arXiv:hep-th/9711200].

\bibitem{holography}
  See, for example, N.~Arkani-Hamed, M.~Porrati and L.~Randall,
  JHEP {\bf 0108}, 017 (2001)
  [arXiv:hep-th/0012148];
  R.~Rattazzi and A.~Zaffaroni,
  JHEP {\bf 0104}, 021 (2001)
  [arXiv:hep-th/0012248].

\bibitem{Goldberger:2007zk}
  W.~D.~Goldberger, B.~Grinstein and W.~Skiba,
  Phys.\ Rev.\ Lett.\  {\bf 100}, 111802 (2008)
  [arXiv:0708.1463 [hep-ph]].

\bibitem{Fan:2008jk}
  J.~Fan, W.~D.~Goldberger, A.~Ross and W.~Skiba,
  Phys.\ Rev.\  D {\bf 79}, 035017 (2009)
  [arXiv:0803.2040 [hep-ph]].

\bibitem{Vecchi:2010gj}
  L.~Vecchi,
  Phys.\ Rev.\  {\bf D82}, 076009 (2010).
  [arXiv:1002.1721 [hep-ph]].

\bibitem{Appelquist:2010gy}
  T.~Appelquist, Y.~Bai,
  Phys.\ Rev.\  {\bf D82}, 071701 (2010).
  [arXiv:1006.4375 [hep-ph]].

\bibitem{LittRad}
  H.~Davoudiasl, T.~McElmurry, A.~Soni,
  Phys.\ Rev.\  {\bf D82}, 115028 (2010).
  [arXiv:1009.0764 [hep-ph]].


\bibitem{WTCM}
  Y.~Bai, M.~Carena and E.~Pont\'on,
  Phys.\ Rev.\ D {\bf 81}, 065004 (2010)
  [arXiv:0809.1658 [hep-ph]].

\bibitem{WTCMrad}
  H.~Davoudiasl and E.~Pont\'on,
  Phys.\ Lett.\ B {\bf 680}, 247 (2009)
  [arXiv:0903.3410 [hep-ph]].

\bibitem{comph1}
  R.~Contino, Y.~Nomura, A.~Pomarol,
  Nucl.\ Phys.\  {\bf B671}, 148-174 (2003)
  [hep-ph/0306259].

\bibitem{comph2}
  K.~Agashe, R.~Contino, A.~Pomarol,
  Nucl.\ Phys.\  {\bf B719}, 165-187 (2005)
  [hep-ph/0412089].


\bibitem{GW1}
  W.~D.~Goldberger and M.~B.~Wise,
  Phys.\ Rev.\ Lett.\  {\bf 83}, 4922 (1999)
  [arXiv:hep-ph/9907447].

\bibitem{GW2}
  W.~D.~Goldberger and M.~B.~Wise,
  Phys.\ Lett.\  B {\bf 475}, 275 (2000)
  [arXiv:hep-ph/9911457].

\bibitem{CHL}
  C.~Cs\'aki, J.~Hubisz and S.~J.~Lee,
  Phys.\ Rev.\  D {\bf 76}, 125015 (2007)
  [arXiv:0705.3844 [hep-ph]].

\bibitem{Early-radion}
For early work on radion physics see, for example:   
C.~Cs\'aki, M.~Graesser, L.~Randall and J.~Terning,
  Phys.\ Rev.\ D {\bf 62}, 045015 (2000)
  [hep-ph/9911406]; 
U.~Mahanta and S.~Rakshit,
  Phys.\ Lett.\ B {\bf 480}, 176 (2000)
  [hep-ph/0002049]; 
G.~F.~Giudice, R.~Rattazzi and J.~D.~Wells,
  Nucl.\ Phys.\ B {\bf 595}, 250 (2001)
  [hep-ph/0002178]; 
S.~Bae, P.~Ko, H.~S.~Lee and J.~Lee,
  Phys.\ Lett.\ B {\bf 487}, 299 (2000)
  [hep-ph/0002224]; 
C.~Cs\'aki, M.~L.~Graesser and G.~D.~Kribs,
  Phys.\ Rev.\  D {\bf 63}, 065002 (2001)
  [arXiv:hep-th/0008151];
J.~L.~Hewett and T.~G.~Rizzo,
  JHEP {\bf 0308}, 028 (2003)
  [hep-ph/0202155]; 
D.~Dominici, B.~Grzadkowski, J.~F.~Gunion and M.~Toharia,
  Nucl.\ Phys.\ B {\bf 671}, 243 (2003)
  [hep-ph/0206192].

\bibitem{Recent-radion}
V.~Barger and M.~Ishida,
  Phys.\ Lett.\ B {\bf 709}, 185 (2012)
  [arXiv:1110.6452 [hep-ph]]; 
H.~de Sandes and R.~Rosenfeld,
  Phys.\ Rev.\ D {\bf 85}, 053003 (2012)
  [arXiv:1111.2006 [hep-ph]]; 
K.~Cheung and T.~-C.~Yuan,
  Phys.\ Rev.\ Lett.\  {\bf 108}, 141602 (2012)
  [arXiv:1112.4146 [hep-ph]]; 
B.~Grzadkowski, J.~F.~Gunion and M.~Toharia,
  Phys.\ Lett.\ B {\bf 712}, 70 (2012)
  [arXiv:1202.5017 [hep-ph]]; 
H.~Kubota and M.~Nojiri,
  arXiv:1207.0621 [hep-ph].  




\bibitem{Peskin:1991sw}
  M.~E.~Peskin and T.~Takeuchi,
  Phys.\ Rev.\  D {\bf 46}, 381 (1992).


\bibitem{Georgi:1992dw}
  H.~Georgi,
  Phys.\ Lett.\  {\bf B298}, 187-189 (1993).
  [hep-ph/9207278].

\bibitem{LRS}
  H.~Davoudiasl, G.~Perez and A.~Soni,
  Phys.\ Lett.\  B {\bf 665}, 67 (2008)
  [arXiv:0802.0203 [hep-ph]].


\bibitem{Bauer:2008xb} 
  M.~Bauer, S.~Casagrande, L.~Gr\"under, U.~Haisch and M.~Neubert,
  Phys.\ Rev.\ D {\bf 79}, 076001 (2009)
  [arXiv:0811.3678 [hep-ph]].


\bibitem{GfitterOb}
GFitter Group, {\tt http://gfitter.desy.de/Oblique\_Parameters/\,.}

\bibitem{Peskin:2001rw}
 M.~E.~Peskin, J.~D.~Wells,
 Phys.\ Rev.\  {\bf D64}, 093003 (2001)
[hep-ph/0101342].

\bibitem{Davoudiasl:2000wi} 
  H.~Davoudiasl, J.~L.~Hewett and T.~G.~Rizzo,
  Phys.\ Rev.\ D {\bf 63}, 075004 (2001)
  [hep-ph/0006041].


\bibitem{M5}
See, for example, the discussion in the fourth paper in Ref.~\cite{LHCKK}.

\bibitem{Konstandin:2010cd}
  T.~Konstandin, G.~Nardini and M.~Quiros,
  Phys.\ Rev.\ D {\bf 82}, 083513 (2010)
  [arXiv:1007.1468 [hep-ph]].

\bibitem{Eshel:2011wz}
  Y.~Eshel, S.~J.~Lee, G.~Perez and Y.~Soreq,
  JHEP {\bf 1110}, 015 (2011)
  [arXiv:1106.6218 [hep-ph]].

\bibitem{Georgi:1977gs} 
  H.~M.~Georgi, S.~L.~Glashow, M.~E.~Machacek and D.~V.~Nanopoulos,
  Phys.\ Rev.\ Lett.\  {\bf 40}, 692 (1978).

\bibitem{HiggsCSWG}
LHC Higgs Cross Section Working Group, {\tt https://twiki.cern.ch/twiki/bin/view/LHCPhysics/}.

\bibitem{Harlander:2002wh} 
  R.~V.~Harlander and W.~B.~Kilgore,
  Phys.\ Rev.\ Lett.\  {\bf 88}, 201801 (2002)
  [hep-ph/0201206].

\bibitem{Catani:2003zt} 
  S.~Catani, D.~de Florian, M.~Grazzini and P.~Nason,
  JHEP {\bf 0307}, 028 (2003)
  [hep-ph/0306211].


\bibitem{Hahn:2004fe} 
  T.~Hahn,
  Comput.\ Phys.\ Commun.\  {\bf 168}, 78 (2005)
  [hep-ph/0404043].

\bibitem{Alwall:2011uj} 
  J.~Alwall, M.~Herquet, F.~Maltoni, O.~Mattelaer and T.~Stelzer,
  JHEP {\bf 1106}, 128 (2011)
  [arXiv:1106.0522 [hep-ph]].

\bibitem{Lai:2010vv} 
  H.-L.~Lai, M.~Guzzi, J.~Huston, Z.~Li, P.~M.~Nadolsky, J.~Pumplin and C.-P.~Yuan,
  Phys.\ Rev.\ D {\bf 82}, 074024 (2010)
  [arXiv:1007.2241 [hep-ph]].

\bibitem{CMS-HWW}
  CMS Collaboration,
  CMS PAS HIG-11-024.

\bibitem{ATLAS-HWW}
  ATLAS Collaboration,
  ATLAS-CONF-2012-012.

\bibitem{Barr:2009mx} 
  A.~J.~Barr, B.~Gripaios and C.~G.~Lester,
  JHEP {\bf 0907}, 072 (2009)
  [arXiv:0902.4864 [hep-ph]].

\bibitem{Barr:2011si} 
  A.~J.~Barr, B.~Gripaios and C.~G.~Lester,
  Phys.\ Lett.\ B {\bf 713}, 495 (2012)
  [arXiv:1110.2452 [hep-ph]].

\bibitem{4bjets} 
  T.~Binoth, N.~Greiner, A.~Guffanti, J.~Reuter, J.-P.~Guillet and T.~Reiter,
  Phys.\ Lett.\ B {\bf 685}, 293 (2010)
  [arXiv:0910.4379 [hep-ph]], 
N.~Greiner, A.~Guffanti, T.~Reiter and J.~Reuter,
  Phys.\ Rev.\ Lett.\  {\bf 107}, 102002 (2011)
  [arXiv:1105.3624 [hep-ph]].

\bibitem{Lafaye:2000ec} 
  R.~Lafaye, D.~J.~Miller, M.~M\"uhlleitner and S.~Moretti,
  hep-ph/0002238.

\bibitem{:2012gk} 
  G.~Aad {\it et al.}  [ATLAS Collaboration],
  Phys.\ Lett.\ B {\bf 716}, 1 (2012)
  [arXiv:1207.7214 [hep-ex]].

\bibitem{:2012gu} 
  S.~Chatrchyan {\it et al.}  [CMS Collaboration],
  Phys.\ Lett.\ B {\bf 716}, 30 (2012)
  [arXiv:1207.7235 [hep-ex]].

\end{thebibliography}
\end{document}